

\font\titlefont = cmr10 scaled\magstep 4
 2
\font\sectionfont = cmr10
\font\littlefont = cmr5 
\font\eightrm = cmr8

\def\ss{\scriptstyle}
\def\sss{\scriptscriptstyle}

\newcount\tcflag
\tcflag = 0  

\ifnum\tcflag = 0 \magnification = 1200 \fi  

\global\baselineskip = 1.2\baselineskip 
\global\parskip = 4pt plus 0.3pt 
\global\abovedisplayskip = 18pt plus3pt minus9pt
\global\belowdisplayskip = 18pt plus3pt minus9pt
\global\abovedisplayshortskip = 6pt plus3pt
\global\belowdisplayshortskip = 6pt plus3pt

\def\barsoff{\overfullrule=0pt}


\def\endignore{}
\def\ignore #1\endignore{} 

\newcount\dflag
\dflag = 0


\def\monthname{\ifcase\month 
\or January \or February \or March \or April \or May \or June%
\or July \or August \or September \or October \or November %
\or December 
\fi}

\newcount\dummy
\newcount\minute  
\newcount\hour
\newcount\localtime
\newcount\localday
\localtime = \time
\localday = \day

\def\advanceclock#1#2{ 
\dummy = #1
\multiply\dummy by 60
\advance\dummy by #2
\advance\localtime by \dummy
\ifnum\localtime > 1440 
\advance\localtime by -1440
\advance\localday by 1
\fi}

\def\settime{{\dummy = \localtime %
\divide\dummy by 60%
\hour = \dummy 
\minute = \localtime%
\multiply\dummy by 60%
\advance\minute by -\dummy 
\ifnum\minute < 10
\xdef\spacer{0} 
\else \xdef\spacer{}
\fi %
\ifnum\hour < 12
\xdef\ampm{a.m.} 
\else
\xdef\ampm{p.m.} 
\advance\hour by -12 %
\fi %
\ifnum\hour = 0 \hour = 12 \fi 
\xdef\timestring{\number\hour : \spacer \number\minute%
\thinspace \ampm}}}



\def\endtitle{}
\def\title#1\endtitle{\vskip.5in\titlefont
\global\baselineskip = 2\baselineskip 
#1\vskip.4in
\baselineskip = 0.5\baselineskip\rm}

\def\endauthors{}
\def\authors#1\endauthors{#1}

\def\endabstract{}
\def\abstract#1\endabstract{\vskip .3in%
\centerline{\sectionfont\bf Abstract}%
\vskip .1in
\noindent#1}

\def\nopageonenumber{\footline={\ifnum\pageno<2\hfil\else
\hss\tenrm\folio\hss\fi}}  

\newcount\nsection 
\newcount\nsubsection 

\def\section#1{\global\advance\nsection by 1
\nsubsection=0
\bigskip\noindent\centerline{\sectionfont \bf \number\nsection.\ #1}
\bigskip\rm\nobreak}

\def\subsection#1{\global\advance\nsubsection by 1
\bigskip\noindent\sectionfont \sl \number\nsection.\number\nsubsection)\
#1\bigskip\rm\nobreak}


\def\appendix#1#2{\bigskip\noindent%
\centerline{\sectionfont \bf Appendix #1.\ #2} 
\bigskip\rm\nobreak} 


\newcount\nref 
\global\nref = 1 

\def\therefs{}


\def\ref#1#2{\xdef #1{[\number\nref]} 
\ifnum\nref = 1\global\xdef\therefs{\item{[\number\nref]} #2\ } 
\else
\global\xdef\oldrefs{\therefs}
\global\xdef\therefs{\oldrefs\vskip.1in\item{[\number\nref]} #2\ }%
\fi%
\global\advance\nref by 1
}

\def\listrefs{\vfill\eject\section{References}\therefs}


\newcount\nfoot 
\global\nfoot = 1 

\def\foot#1#2{\xdef #1{(\number\nfoot)} 
\hskip -0.2cm ${}^{\number\nfoot}$
\footnote{}{\vbox{\baselineskip=10pt
\eightrm \hskip -1cm ${}^{\number\nfoot}$ #2}}
\global\advance\nfoot by 1
}


\newcount\nfig 
\global\nfig = 1
\def\thefigs{} 

\def\figure#1#2{\xdef #1{(\number\nfig)}
\ifnum\nfig = 1\global\xdef\thefigs{\item{(\number\nfig)} #2\ }
\else
\global\xdef\oldfigs{\thefigs}
\global\xdef\thefigs{\oldfigs\vskip.1in\item{(\number\nfig)} #2\ }%
\fi%
\global\advance\nfig by 1 } 

\def\fig#1{\xdef #1{(\number\nfig)}
\global\advance\nfig by 1 } 


\newcount\ntab
\global\ntab = 1

\def\table#1{\xdef #1{\number\ntab}
\global\advance\ntab by 1 } 


\newcount\cflag
\newcount\nequation
\global\nequation = 1
\def\eqlabel{(1)}

\def\nexteqno{\ifnum\cflag = 0
\global\advance\nequation by 1
\fi
\global\cflag = 0
\xdef\eqlabel{(\number\nequation)}}

\def\lasteqno{\global\advance\nequation by -1
\xdef\eqlabel{(\number\nequation)}}

\def\label#1{\xdef #1{(\number\nequation)}
\ifnum\dflag = 1
{\escapechar = -1
\xdef\draftname{\littlefont\string#1}}
\fi}

\def\clabel#1#2{\xdef\eqlabel{(\number\nequation #2)}
\global\cflag = 1
\xdef #1{\eqlabel}
\ifnum\dflag = 1
{\escapechar = -1
\xdef\draftname{\string#1}}
\fi}

\def\cclabel#1#2{\xdef\eqlabel{#2)}
\global\cflag = 1
\xdef #1{\eqlabel}
\ifnum\dflag = 1
{\escapechar = -1
\xdef\draftname{\string#1}}
\fi}


\def\eeq{}

\def\eqnn #1\eeq{$$ #1 $$}

\def\eq #1\eeq{
\ifnum\dflag = 0
{\xdef\draftname{\ }}
\fi 
$$ #1
\eqno{\eqlabel \rlap{\ \draftname}} $$
\nexteqno}







\def\eqa #1\eeq{
\ifnum\dflag = 0
{\xdef\draftname{\ }}
\fi 
$$ \eqalignno{ #1 } $$
\global\cflag = 0}


\def\ie{{\it i.e.\/}}

\def\etc{{\it etc.\/}}


\def\anp#1#2#3{{\it Ann.\ Phys. (NY)} {\bf #1} (19#2) #3}

\def\cmp#1#2#3{{\it Comm.\ Math.\ Phys.} {\bf #1} (19#2) #3}

\def\npb#1#2#3{{\it Nucl.\ Phys.} {\bf B#1} (19#2) #3}
\def\plb#1#2#3{{\it Phys.\ Lett.} {\bf #1B} (19#2) #3}

\def\prd#1#2#3{{\it Phys.\ Rev.} {\bf D#1} (19#2) #3}


\global\nulldelimiterspace = 0pt



\def\frac#1#2{{{#1} \over {#2}}\,}  
\def\hf{{1\over 2}}

\def\stack#1#2{\buildrel{#1}\over{#2}}

\def\Square{{\vbox {\hrule height 0.6pt\hbox{\vrule width 0.6pt\hskip 3pt
        \vbox{\vskip 6pt}\hskip 3pt \vrule width 0.6pt}\hrule height 0.6pt}}}
\def\Asl{\hbox{/\kern-.7500em\it A}} 
\def\Dsl{\hbox{/\kern-.6700em\it D}} 
\def\dsl{\hbox{/\kern-.5300em$\partial$}}
\def\pxpsl{\hbox{/\kern-.5600em$p$}}
\def\sslsh{\hbox{/\kern-.5300em$s$}}
\def\epssl{\hbox{/\kern-.5100em$\epsilon$}}
\def\delsl{\hbox{/\kern-.6300em$\nabla$}}
\def\lxpsl{\hbox{/\kern-.4300em$l$}}
\def\elxpsl{\hbox{/\kern-.4500em$\ell$}}
\def\kxpsl{\hbox{/\kern-.5100em$k$}}
\def\qxpsl{\hbox{/\kern-.5000em$q$}}
\def\sla#1{\raise.15ex\hbox{$/$}\kern-.57em #1}



\def\twi{\widetilde}

\def\roughly#1{\mathrel{\raise.3ex\hbox{$#1$\kern-.75em\lower1ex\hbox{$\sim$}}}}

\def\ol#1{\overline{#1}}





\def\Scb{{\cal B}}

\def\Scd{{\cal D}}
\def\Sce{{\cal E}}
\def\Scf{{\cal F}}

\def\Scj{{\cal J}}

\def\Scl{{\cal L}}

\def\Scq{{\cal Q}}
\def\Scr{{\cal R}}


\def\ssa{{\sss A}}
\def\ssb{{\sss B}}

\def\ssl{{\sss L}}

\def\ssq{{\sss Q}}

\def\sst{{\sss T}}

\def\ssv{{\sss V}}


\def\pmb#1{\setbox0=\hbox{#1}%
\kern-.025em\copy0\kern-\wd0
\kern.05em\copy0\kern-\wd0
\kern-.025em\raise.0433em\box0}


\font\jlgtenbrm=cmbx10
\font\jlgtenbit=cmmib10
\font\jlgtenbsy=cmbsy10
\font\jlgsevenbrm=cmbx10 at 7pt
\font\jlgsevenbsy=cmbsy10 at 7pt
\font\jlgsevenbit=cmmib10 at 7pt
\font\jlgfivebrm=cmbx10 at 5pt
\font\jlgfivebsy=cmbsy10 at 5pt
\font\jlgfivebit=cmmib10 at 5pt
\newfam\jlgbrm

\textfont\jlgbrm=\jlgtenbrm
\scriptfont\jlgbrm=\jlgsevenbrm
\scriptscriptfont\jlgbrm=\jlgfivebrm
\newfam\jlgbit

\textfont\jlgbit=\jlgtenbit
\scriptfont\jlgbit=\jlgsevenbit
\scriptscriptfont\jlgbit=\jlgfivebit
\newfam\jlgbsy

\textfont\jlgbsy=\jlgtenbsy
\scriptfont\jlgbsy=\jlgsevenbsy
\scriptscriptfont\jlgbsy=\jlgfivebsy
\newcount\jlgcode
\newcount\jlgfam
\newcount\jlgchar
\newcount\jlgtmp
\def\bolded#1{
        \jlgcode\the#1 \divide\jlgcode by 4096
        \jlgtmp\the\jlgcode \multiply\jlgtmp by 4096
        \jlgfam\the#1 \advance\jlgfam by -\the\jlgtmp
        \divide\jlgfam by 256
        \jlgtmp\the\jlgcode \multiply\jlgtmp by 16
	\advance\jlgtmp by \the\jlgfam
	\multiply\jlgtmp by 256
        \jlgchar\the#1 \advance\jlgchar by -\the\jlgtmp
        \advance\jlgfam by \the\jlgbrm
        \jlgtmp\the\jlgcode
        \multiply\jlgtmp by 16
        \advance\jlgtmp by \the\jlgfam
        \multiply\jlgtmp by 256
        \advance\jlgtmp by \the\jlgchar
        \mathchar\the\jlgtmp
}


\def\det{\mathop{\rm det}}


\def\bra#1{\langle #1 |}
\def\ket#1{| #1 \rangle}

\def\Avg#1{\left\langle #1 \right\rangle}



\def\vacbra{{\bra 0}}
\def\vac{{\ket 0}}





\nopageonenumber
\baselineskip = 18pt
\barsoff


\def\bk{\item{}}

\def\twtw{(2,2)}
\def\onon{(1,1)}

\def\veps{\varepsilon}
\def\mn{{\mu\nu}}

\def\twch{twisted-chiral\ }

\def\sdet{{\rm sdet}\ }

\def\SC{{\sss SC}}
\def\SL{{\sss SL}}
\def\FP{{\sss FP}}
\def\LM{{\sss LM}}
\def\FPD{Fadeev-Popov-DeWitt}

\def\beq{\eq}

\def\L{\Lambda}


\line{hep-th/9809085  \hfil McGill-98/18, BRX-TH-438, Mexico-IFUNAM-FT98-10. }

\vskip .05in
\title
\centerline{Spacetime Duality and  }
\centerline{Superduality}
\endtitle

\vskip 0.1in
\authors
\centerline{C.P. Burgess,${}^a$ M. T. Grisaru,${}^b$
M. Kamela,${}^a$ M. E. Knutt-Wehlau,${}^a$}
\centerline{P. Page${}^a$,
F. Quevedo,${}^c$\footnote{*}{\vbox{\baselineskip=10pt
\eightrm John Simon Guggenheim Fellow.
Address after August 15, 1998: D.A.M.T.P., Cambridge University,
Cambridge, UK.}} and M. Zebarjad${}^d$}
\vskip .05in
\centerline{\it ${}^a$ Physics Department, McGill University}
\centerline{\it 3600 University St., Montr\'eal, Qu\'ebec, Canada,
H3A 2T8.}
\vskip .05in
\centerline{\it ${}^b$ Physics Department, Brandeis University}
\centerline{\it Waltham, Massachusetts, 02454 USA.}
\vskip .05in
\centerline{\it ${}^c$ Instituto de F\'isica, Universidad Nacional
Aut\'onoma de M\'exico}
\centerline{\it Apartado Postal 20-364, 01000  M\'exico D.F.,
M\'exico.}
\vskip .05in
\centerline{\it ${}^d$ Physics Department and Biruni Observatory,
Shiraz University, Shiraz 71454, Iran.
}
\endauthors

\abstract
\vbox{\baselineskip 15pt
We introduce a new class of duality symmetries amongst quantum field
theories. The new class is based upon global {\it spacetime}
symmetries, such as Poincar\'e invariance and supersymmetry,
in the same way as the existing duality transformations are
based on global {\it internal} symmetries.
We illustrate these new
duality transformations by dualizing several scalar and spin-half
field theories in 1+1 spacetime dimensions, involving
nonsupersymmetric as well as $(1,1)$ and $(2,2)$
supersymmetric models. For $(2,2)$ models the new duality
transformations can interchange  chiral and twisted chiral  multiplets.}
\endabstract
%
\vfill\eject

\section{Introduction}

\ref\stringdual{For a recent review see A. Sen, hep-th/9802051.}

\ref\maldacenarefs{J. Maldacena, preprint HUTP-98-A097, hep-th/9711200.}

\ref\bosorefs{C.P. Burgess and F. Quevedo,
\npb{421}{94}{373};\bk \plb{329}{94}{4}.}

\ref\formrefs{For a review with extensive references to the
literature see, for instance, F. Quevedo,
{\it Nucl. Phys.} (Proc. Suppl.) {\bf 61A } (1998), hep-th/9706210.}

Duality symmetries are dramatically changing our
understanding of quantum field theories
 by demonstrating the equivalence
of many different models which previously had been
thought to be entirely distinct. This is having enormous
implications for our understanding of string theory,
with more and more string models being discovered
to be duals of one another \stringdual. At this writing evidence
is accumulating in support of a duality between
some types of string theories and four-dimensional gauge
theories \maldacenarefs. Duality also has more
prosaic implications, implying in two dimensions
the equivalence between fermions and bosons \bosorefs,
and in $D$ dimensions the equivalence between massless
antisymmetric tensor fields of rank $r$ and
rank $D-r-2$ \formrefs, and so on.

\ref\buscher{T. Buscher, \plb{194}{87}{59}; \bk
\plb{201}{88}{466}.}
\ref\rocekverlinde{M. Ro\v cek and E. Verlinde, \npb{373}{92}{630}.}

\ref\earlydual{See for instance, N.J. Hitchin, A. Karlhede,
U. Lindstr\"{o}m and M. Ro\v{c}ek, \cmp{108}{87}{535} and
references therein.}

In its most concrete form, duality may be considered
to be the following algorithm for constructing
an alternate description of a given field theory, $\Scf$,
provided that $\Scf$ has a global internal symmetry group, $G$
\buscher, \rocekverlinde, \earlydual.
The duality algorithm instructs us to gauge the symmetry $G$, but
also to impose a constraint that eliminates the corresponding
gauge field strength. This constraint is to be implemented
by introducing a Lagrange multiplier field $\L$. Thus,
by construction, integrating over $\L$ and fixing a gauge
for $G$ reproduces the original model, $\Scf$. The dual,
$\twi\Scf$, is found by changing the order of
functional integration, leaving for last the field $\L$,
which is the fundamental field of the dual theory.

\ref\yau{A. Strominger, S.-T. Yau and E. Zaslow, \npb{479}{96}{243}.}

\ref\plesser{D.R. Morrison and M.R. Plesser,
{\it Nucl. Phys.} (Proc. Suppl.) {\bf 46 } (1996),
hep-th/9508107.}

As our understanding improves, many of the
equivalences among field theories prove to be special
cases of this duality algorithm. Some equivalences --- such as
fermionization (as opposed to bosonization), or the
string-theoretic equivalence known as
mirror symmetry --- have resisted such an understanding
however, and at present do not appear to be based on
 any pre-existing internal global
symmetries. For mirror symmetry
the situation is even worse. Although there is strong evidence
in its favour,  an explicit proof of the validity of mirror symmetry
 still eludes us,  interesting insights \yau\  notwithstanding.
Despite the failure of recent valiant efforts \plesser,
one suspects that  a better understanding of such symmetries
in terms of the duality algorithm must shed light on their origins
and domains of validity.

Motivated by these considerations, it is interesting
to try to extend the  duality formalism beyond its
present limits and to explore its broadest consequences.
In this article we make a first step
towards a generalization of the duality formalism.
Up until now only global internal symmetries
have been used to dualize a theory, even though the quantum field
theories of interest also have
spacetime symmetries, such as Poincar\'e invariance or
supersymmetry. Our purpose
here is to propose a new class of duality transformations
which are based on these spacetime
symmetries.
Besides its intrinsic formal interest
as a more general way to dualize arbitrary field theories, we expect
that our extension may be of use in better understanding
some of the open questions mentioned above.

When duality is based on a spacetime symmetry, the
gauging means coupling the original model to gravity
(or supergravity). The constraint which removes the
gauge degrees of freedom now imposes flat spacetime
(or a vanishing gravitino, or both). As in the usual construction,
it is the interchange of the order of functional integrations
which gives rise to the dual theory.

In principle this very general prescription
provides a concrete construction of a dual theory
for {\it any} relativistic field theory, with potentially
far-reaching consequences. In particular, global
internal symmetries are not required; our procedure  should
be applicable, for example, to string backgrounds without
isometries, such as Calabi-Yau compactifications
and conformal field theories based on cosets $G/H$, having a
non-Abelian group $H$.

We remark in passing that our new duality transformation
should not be confused with the well known $T$-duality
of strings propagating in curved spacetimes. For $T$-duality
the original theory is the 2D worldsheet sigma model and
the metric of the curved `target' space appears as a coupling
of the worldsheet fields. Even though $T$-duality is based
on the existence of isometries of the target metric, within
the 2D worldsheet field-theory context this isometry
is an ordinary internal symmetry. The duality we propose,
on the other hand, is based on 2D Lorentz
invariance or supersymmetry on the worldsheet itself.

We present our ideas as follows: In the next section, \S2,
we review one of the simplest duality transformations,
based on the symmetry $X \to X + \hbox{\rm
constant}$ for a scalar field $X$, as a paradigm of the steps we
follow for the spacetime symmetries. \S3\ through \S6\ then
present spacetime duality through a series of simple
examples, all involving massless fields in 1+1 dimensions.
\S3\ starts with the simplest possible case, the
dualization of a scalar field coupled to a background
gravitational field in 1+1 dimensions. In \S4\ we generalize
the result of \S3\ to find the dual of a Dirac
fermion in 1+1 dimensions. This turns out to provide an
alternative bosonization method  that differs
in detail from the standard one. In \S5\ we
discuss superduality, which is the extension of our
formalism to dualization using supersymmetry
itself as the initial global symmetry. Again we
illustrate the method using the simplest case of
a $(1,1)$ supersymmetric Wess-Zumino model, although
the dual theory is in this case equivalent to the
original one. \S6\ examines the same construction
for a $(2,2)$ supersymmetric Wess-Zumino model, where
we find that duality maps chiral and twisted chiral
multiplets into one another.  In
\S7\  we summarize our results.  In the appendix
we discuss some
ambiguities in the expression for the (2,2) component
superconformal anomaly.

\section{The Duality Algorithm}

In order to better describe the technical details of
spacetime duality, we pause here to first discuss ordinary
duality in its simplest setting. Consider therefore
a free massless real scalar field in 1+1 dimensions,
with action $S=-\hf \, \int\partial_\mu X
\partial^\mu X$. We compute the generating functional
for this model in the presence of an external background
gauge field, $a_\mu$:
\beq
Z[a_\mu]=\int [\Scd X] \exp  \left\{ -\, \frac{i}{2}\int{d^2x\,
\,  \left(\partial_\mu  X -a_\mu\right)
\left( \partial^\mu X -a^\mu\right) } \right\} .
\label\scalaro
\eeq
The generating functional defined by \scalaro\ is invariant
under the background gauge symmetry:
$a_\mu \to a_\mu + \partial_\mu \omega$,
for arbitrary $\omega$, as may be seen by
performing the change of integration variable:
$X\rightarrow X + \omega$.

We include the coupling to $a_\mu$ in order to have
an argument on which the result depends after performing
the path integral. This is important since in what follows
we ignore overall constants throughout
when performing functional
integrals. Physically, differentiation of $Z$ with
respect to $a^\mu$ gives the
correlation
functions for the operator $\partial_\mu X$, which
is also the Noether current of the symmetry for which
$a_\mu$ is the gauge potential.

When considering the
dependence on background fields, such as $a_\mu$,
it is important to keep in mind that some dependence may appear
implicitly in the definition of the  functional
integral measure, particularly for the background gravitational
fields we encounter in subsequent sections.
This complication does not arise for the simple system considered
here.

In order to dualize this system we follow the following
steps:

\item{1.}
Gauge the background symmetry $X\rightarrow
X + \omega$ by introducing a
dynamical gauge field $A_\mu$ (\ie\ one over which a functional
integral is to be performed).
\item{2.}
Choose a gauge-fixing condition, $f=0$,
(together with the corresponding \FPD\ determinant, $J_\FP$),
as is required to evaluate the path integral over $A_\mu$.
\item{3.}
Impose a gauge-covariant constraint which implies $A_\mu$
is pure gauge. This constraint, together with the gauge condition
just described, is designed to ensure that the path
integral over $A_\mu$ is equivalent to evaluating the integrand
at the configuration $A_\mu = 0$.
\item{4.}
Rewrite the constraint of item 3 by introducing a Lagrange multiplier
($\L$) whose path integration imposes this constraint. Integrating
over $\L$, and then over $A_\mu$, therefore reproduces the original
theory, eq.~\scalaro.
\item{5.}
Finally, perform the path integral in a different order:
integrate first over $X$ and $A_\mu$, leaving the
integral over $\L$ unperformed. The result is
the `dual' theory, with $\L$ playing the role of the
dual field variable.

For the present example, steps 1 through 4 amount to
rewriting eq.~\scalaro\ in the following way:
\eq
\eqalign{
Z[a_\mu] & = \int [\Scd X] \,
[\Scd A_\mu] \, [\Scd   \L] \; \Delta[f] \;  J_\FP \times \cr
&  \qquad \exp \left\{ -\,\frac{i}{2}\int d^2x\,  \Bigl[
\left(\partial_\mu X -a_\mu-A_\mu\right)
\left(  \partial^\mu X -a^\mu-A^\mu\right)  +2 \L \;
\varepsilon^{\mu\nu}\partial_\mu
A_\nu \Bigr] \right\} . \cr}
\label\scalartt
\eeq
Here $\Delta[f]$ is the functional delta function which
imposes the gauge condition $f=0$, and $J_\FP$ is the
associated \FPD\ determinant.

\ref\topcomprefs{
See reference \bosorefs, as well as F. Quevedo and
 C. Trugenberger, \npb{501}{97}{143}.}

It is clear that integrating over $\L$ gives a functional
delta function which imposes the constraint
$\varepsilon^{\mu\nu}\partial_\mu A_\nu=0$, \ie\
$A_\mu$ has vanishing field strength. Using the gauge fixing
condition $f \equiv \partial^\mu A_\mu = 0$ and ignoring
overall $a_\mu$-independent constants,
the integration over $A_\mu$, barring topological complications,
\foot\topcomp{See, however,
refs.~\topcomprefs\ for a discussion of duality on spaces
with nontrivial topology.}  is accomplished
  by simply
setting $A_\mu = 0$ everywhere, thus
recovering the original
generating functional \scalaro.

 On the other
hand, first integrating $X$ and $A_\mu$,
 it is more convenient to use the gauge $f \equiv X = 0$
to perform the $X$ integration.
Then the remaining
integral over $A_\mu$ is Gaussian, giving the dual result:
\beq
Z[a_\mu] = \int [\Scd \L] \exp  \left\{-\, \frac{i}{2} \int{d^2x\,
\left[  \partial_\mu \L \partial^\mu \L - 2
\varepsilon^{\mu\nu} a_\mu\partial_\nu \L   \right] } \right\} .
\label\scalaroo
\eeq

The significance of the dual formulation lies in the
observation that the coupling to $a_\mu$ differs in
eq.~\scalaroo\ from that of eq.~\scalaro. In particular,
the difference in the term linear in $a_\mu$ in the respective
actions indicates that the field operators dualize
according to the standard relation:
\eq
\partial_\mu X \leftrightarrow \veps_{\mu\nu}
 \, \partial^\nu \L .
\label\wkdr
\eeq

In addition, notice that the action for $X$ contains
the quadratic term, $a_\mu a^\mu$, but no such
term appears in the dual action for $\L$.  This also
has physical implications, since twice differentiating
eqs.~\scalaro\ and \scalaroo\ implies:
\eq
(-i)^2  \left.{\delta^2 Z \over \delta a^\mu(x) a^\nu(y)}\right|_{a_\mu=0}
= \Avg{\veps_{\mu\alpha}\partial^\alpha \L(x)
\; \veps_{\nu\beta} \partial^\beta \L(y)}
= \Avg{\partial_\mu X(x) \; \partial_\nu X(y)}
  +i \eta_{\mu\nu} \, \delta^2(x-y) ,
\label\cterm
\eeq
where $\eta_{\mu\nu}$ is the usual Minkowski-space metric.
$\Avg{\cdots}$ here indicates the covariant $T^*$
product, which is related to the garden-variety
time-ordered ($T$) product by, for example,
$\vacbra T^*[\partial_\mu X(x)
\partial_\nu X(y)] \vac \equiv \partial_\mu \partial_\nu
\vacbra T[X(x) X(y)] \vac$.

The $\delta$-function contact term in the last of the
equalities in eq.~\cterm\ is just what is required for this equation to
make sense. After all, the correspondence, eq.~\wkdr,
implies that time derivatives, $\partial_t X$, dualize
to space derivatives, $\partial_x \L$, and while
time derivatives get $\delta$-function contributions
when the derivatives hit the time ordering, space
derivatives do not. The contact term of eq.~\cterm\
is just what is required to make both sides of eq.~\cterm\
agree.

\section{Spacetime Duality I: The Scalar Field}

We now extend the duality algorithm
to spacetime symmetries. To describe the procedure
we focus first on the simplest case, that of a massless
real scalar field in 1+1 dimensions. Although this
model proves to be self-dual under the construction
we outline, the same does not prove to be true
for some of the models considered
in subsequent sections.

As in the previous example, we must choose a
background field  as the argument of the
generating functional. In this
and later examples we choose it to be
a background gravitational field  $h_{\mu\nu}$.
With this choice the
action is
\eq
S[h,X] = -\frac{1}{2} \int d^2 x \sqrt{-h}
h^{\mu \nu}  \partial_\mu X \partial_\nu X
= \frac{1}{2} \int d^2 x \sqrt{-h}   X  \Square_h X,
\label\curvescalar
\eeq
where $\Square_h =  h^{\mu\nu} \nabla_\mu \nabla_\nu
= \frac{1}{\sqrt{-h}}  \partial_\mu (\sqrt{-h} \, h^{\mu \nu}
\partial_\nu)$, and the suffix $h$ emphasizes its
dependence on the metric $h_{\mu\nu}$.
We denote the same quantity without subscripts
 built using the Minkowski-space
metric $\eta_{\mu\nu}$ by
$\Square = \eta^{\mu \nu}\partial_{\mu} \partial_{\nu}$.

\ref\polyakovei{A. M. Polyakov, \plb{103}{81}{207}.}
\ref\nakahara{M. Nakahara, {\it Geometry, Topology and
Physics}, 1991.}
\ref\fujikawaeth{K. Fujikawa, \npb{226}{83}{437}.}
\ref\fujikawaetalee{K. Fujikawa, U. Lindstr\"{o}m, N.K.
Nielsen, M. Ro\v{c}ek and  P. van Nieuwenhuizen, \prd{37}{88}{391}.}

The functional integral on which we perform the duality
algorithm is
\beq
Z[h_{\mu\nu}]=\int\, [\Scd X]_h \, e^{i S[h, X]} \, .
\label\zscalar
\eeq
The subscript `$h$' here is a reminder that the measure,
$[\Scd X]_h$, depends implicitly on the field $h_{\mu\nu}$.
This dependence can be found explicitly using any of
a number of methods for defining a
measure \polyakovei, \nakahara, \fujikawaeth,
\fujikawaetalee\ invariant under general coordinate
transformations (GCTs).
Any such measure is subject to a conformal anomaly,
and since any metric in 1+1 dimensions is conformally flat,
the conformal anomaly may be used to explicitly
display the metric dependence of $[\Scd X]_h$. Thus,
adopting coordinates for which $h_{\mu\nu} =
e^\varphi \, \eta_{\mu\nu}$, we have $[\Scd X]_h
= [\Scd X]_\eta \; \exp(i S_\ssl[\eta,\varphi])$,
where generally  $S_\ssl$ denotes the Liouville action
\eq
\eqalign{
S_\ssl [h,\phi] &\equiv  -\; \frac{1}{48\pi} \int d^2x
\sqrt{-h}  \left( - \; \frac{1}{2} h^{\mu\nu}
\partial_\mu \phi\partial_\nu \phi + \Scr_h
\phi + \mu^2 e^\phi\right) \cr
&= -\, \frac{1}{96  \pi} \left\{  \int d^2 x
\; \sqrt{-g} \; \left( {\cal R}_g
\, \frac{1}{\Square_g} \, {\cal R}_{g} + \mu^2 \right)
\right. \cr
&
\qquad\qquad \qquad\qquad \left.
-  \int d^2 x  \; \sqrt{-h}  \; \left( {\cal R}_h
\, \frac{1}{\Square_h} \, {\cal R}_{h} \right) \right\} . \cr}
\label\liouville
\eeq
Here ${\cal R}_h$ is the curvature scalar defined from the
metric $h_{\mu\nu}$, and $\Scr_g$ is the same quantity
constructed from the conformally-related metric
$g_\mn = e^\phi \, h_\mn$.  We adjust the
regularization-dependent
scale  $\mu$ such as to cancel any such term appearing
elsewhere in the path integral, and so  we omit it in what follows.
As usual, $\Square_h^{-1} \; \Scr_h$ denotes
the convolution $\int d^2y \; \sqrt{-h} \; G_h(x,y) \, \Scr_h(y)$,
of $\Scr_h$ with the Feynman Green's function for $\Square_h$:
$\Square_h \, G_h(x,y) = \delta^2(x-y)/\sqrt{-h}$.

We note that the conformal invariance of the action,
eq.~\curvescalar, permits the functional integral over $X$
in  eq.~\zscalar\ to be explicitly evaluated, yielding
\eq
\eqalign{
[\det(-\Square_h)]^{-1/2} &=
Z[h_{\mu\nu}] = Z[e^\varphi \, \eta_{\mu\nu}]
= Z[\eta_{\mu\nu}] \; e^{i S_\ssl[\eta,\varphi]} =
[\det (-\Square)]^{-1/2}  \;  e^{i S_\ssl[\eta,\varphi]} \cr
&=  [\det (-\Square)]^{-1/2} \exp \left\{ -\; \frac{i}{96
\pi} \int d^2 x  \; \sqrt{-h}\;
\left( {\cal R}_h \, \frac{1}{\Square_h} \,
{\cal R}_{h} \right) \right\}. \cr}
\label\xresult
\eeq

With these preliminaries in hand, we now turn to the
dualization of this model, following the steps outlined
in \S2. We base the duality on the spacetime symmetries
of the model, which we gauge by coupling the system
to a {\it dynamical} gravitational field, $g_{\mu\nu}$,
over which we must functionally integrate.
To this we also add a generally-covariant
constraint which forces the dynamical
field $g_{\mu\nu}$ to be gauge-equivalent to
the background metric $h_{\mu\nu}$, thereby
making the gauged system identical to the
original, eq.~\zscalar.  Duality is then achieved by
interchanging the order of functional integrations.

\ref\CGFPref{
A. M. Polyakov,  \plb{103}{81}{207}; \bk
 {\it Gauge Fields and
Strings}, Harwood, 1987.
}

Therefore we start with the following gauged functional
integral
\eq
Z[h_{\mu\nu}] = \int [\Scd X]_g  [\Scd g_{\mu\nu}]_g
\, \Delta_\LM(g_{\mu\nu}, h_{\mu\nu}) \,
e^{iS[g,X]} ,
\label\scalar
\eeq
where $\Delta_\LM(g_{\mu\nu}, h_{\mu\nu})$ implements
the constraint which forces $g_{\mu\nu}$ to agree
with $h_{\mu\nu}$. (We will have more to say
about this constraint shortly.) To define the measure $[\Scd g_{\mu\nu}]$,
we write the dynamical metric as a combined coordinate and
conformal transformation of the background,
$g_{\mu\nu} = (e^{\phi} h_{\mu\nu})^{\xi}$, and so
write $[\Scd g_\mn]$ as $[\Scd \phi ]_g [\Scd \xi^\mu]_g \Delta[f]
\, [J_\FP]_g$.
 Here $f^\mu = 0$ denotes the coordinate condition
which fixes $\xi^\mu$, and for which we choose
conformal gauge (\ie\ we choose coordinates
so that $g_{\mu\nu} = e^\phi \, h_{\mu\nu}$).
With this choice the \FPD\
determinant becomes $[J_\FP]_g = \det[- \Square^v_g
+ \frac{1}{2} \Scr_g]^{1/2}$, where $\Square^v_g$ is
the Laplacian operating on vector fields. (For later
reference we record the contribution of this determinant
to the conformal anomaly, $[J_\FP]_{e^\phi h}
= [J_\FP]_h \, e^{-26 i S_\ssl[h_\mn,\phi]}$.)
Combining the above definitions permits eq.~\scalar\ to be written
in the following way
\eq
Z[h_{\mu\nu}] = \int [\Scd X]_{e^{\phi } h}  \,
[\Scd \phi ]_{e^{\phi } h} \, [J_\FP]_ {e^{\phi } h} \,
\, \Delta_\LM( {e^{\phi } h_{\mu\nu}} , h_{\mu\nu})
\;  e^{iS[ {e^{\phi } h_{\mu\nu}}, X ]} .
\label\scalartwo
\eeq

At this point we turn to the construction of a suitable
constraint term, $\Delta_\LM(g_\mn,h_\mn)$. We are
guided in this construction by two requirements. First,
it must be proportional to $\Delta[\phi ]$, in order to remove
 the integration over $\phi$  in
eq.~\scalartwo\  by setting $\phi = 0$. Second, it must remove
the $\phi$-independent factor $[J_\FP]_h$ in this
equation, since this does not appear in the original
expression eq.~\zscalar\ for $Z[h_\mn]$. These two conditions
do not suffice to fix $\Delta_\LM$ completely, since
they leave the freedom to multiply by an arbitrary
function which approaches unity as $\phi \to 0$. We
use this freedom to combine as many factors of $h_\mn$
and $\phi$ together into $g_\mn$'s as possible, leading
to the following choice
\eq
\Delta_\LM[ g_{\mu\nu} , h_{\mu\nu}]
=\int [\Scd \L]_g  \, \exp\left\{ - i \int d^2 x
\Bigl( \sqrt{-g} \, {\cal R}_g - \sqrt{-h}\,  {\cal R}_h
\Bigr) \L \right\} \; { \det (-\Square_g) \over [J_\FP]_g } ,
\label\deltaLMcov
\eeq
so that
\eq
\eqalign{
[J_\FP]_{e^\phi \, h} \; \Delta_\LM(e^\phi h_\mn, h_\mn)
&=  \det (-\Square_{e^\phi  h}) \;
\int [\Scd \Lambda]_{e^\phi  h}  \, \exp\left\{ - i \int d^2 x
\sqrt{-h} \Bigl( \Lambda \, \Square_h \phi  \Bigr)      \right\} \cr
&=  \det (-\Square_h) \; \int [\Scd \Lambda]_h \, \exp\left\{ - i \int d^2 x
\sqrt{-h} \Bigl( \Lambda \, \Square_h \phi  \Bigr)      \right\}
\; e^{-i S_\ssl[h,\phi]} \cr
&= \Delta[\phi]  \; e^{-i S_\ssl[h, \phi]} ~~.\cr }
\label\deltaLMfva
\eeq
These manipulations use the conformal anomaly,
$[\Scd \L]_{e^\phi h} = [\Scd \L]_h \, e^{i S_\ssl[h_\mn,\phi]}$,
the relation between curvatures for conformally related metrics
in two dimensions
$ \sqrt{-g} \, {\cal R}_g - \sqrt{-h}\,  {\cal R}_h= \sqrt{-h}\, \Square_h
\phi$,
as well as eq.~\xresult. In the third equality,
the quantity $\Delta[\phi]$ is the functional delta function,
and this equation follows from the previous lines
after performing the change of variables $\L \to -
\Square_h \, \L$.

To verify that eq.~\scalartwo\ is equivalent to our
starting point eq.~\zscalar, we
insert the last of eqs.~\deltaLMfva\ into eq.~\scalartwo\
and use the functional delta function
to perform the $\phi$ path integral, thus setting $\phi =0$
everywhere in the integrand.
  What remains is
precisely eq.~\zscalar.

In order to obtain the dual we again use eqs.~\deltaLMfva\
for $\Delta_\LM$ in eq.~\scalartwo, but this time perform
the integrals over $X$ and $\phi$. We have
\eq
\eqalign{
Z[h_{\mu\nu}] &= \det(-\Square_h) \,
\int[\Scd X]_h \, [\Scd \phi ]_h \,
[\Scd \L]_h \,  \exp \;  \left\{ i \int d^2  x \sqrt{-h}
\; \left[ \frac12 \; X \Square_h X \right. \right. \cr
& \qquad\qquad \qquad\qquad
\left.\left. -\;  {1 \over 48 \pi} \left( \hf \, \phi
\Square_h \phi + \Scr_h \, \phi \right) -
\L \, \Square_h \, \phi \right] \right\} ,\cr
&= \det(-\Square_h) \,
\int[\Scd X]_h \, [\Scd \phi ]_h \,
[\Scd \L]_h \,  \exp \;  \left\{ \frac{i}{2} \int d^2
x \sqrt{-h} \, \left( X \Square_h X  \phantom{\frac12}
\right. \right. \cr
&  \qquad\qquad \qquad\qquad
-\;  {1 \over 48 \pi} \left[ \left( \phi  +
 \frac{1}{\Square_h} \, {\cal R}_h  +
48 \pi \,  \L \right) \Square_h
 \left( \phi  + \frac{1}{\Square_h} {\cal R}_h
+ 48 \pi \,  \L \right) \right.  \cr
&  \qquad\qquad \qquad\qquad
\left. \left. \left. + \, \frac{1}{48 \pi} \, \left(
\frac{1}{\Square_h} \, {\cal R}_h  +  48 \pi\, \L
\right) \Square_h \left( \frac{1}{\Square_h} \,
{\cal R}_h +  48 \pi\, \L
\right)\right] \right) \right\},\cr}
\label\cthesq
\eeq
where the last equality is obtained from the
first by completing the square.
Finally, we perform
the remaining two Gaussian integrations over
$X$ and over $\phi $, producing thereby two factors of
$[\det(-\Square_h)]^{-1/2}$ which precisely cancel
the determinant which appears on the right-hand-side
 of eq.~\cthesq. After rescaling
$\L \to \L/\sqrt{48 \pi}$, we are left with the dual
expression for $Z[h_\mn]$
\eq
\eqalign{
Z[h_{\mu \nu}] &=  \int [\Scd \L]_h \,  \exp  \left\{
\frac{i}{2} \int d^2x \sqrt{-h} \left(  \L  + \frac{1}{\sqrt{48 \pi}}
\, \frac{1}{ {\Square}_h}  \, {\cal{R}}_h \right)
{\Square}_h \left(  {\Lambda}  + \frac{1}{\sqrt{48 \pi}}
\, \frac{1}{ {\Square}_h}  \, {\cal{R}}_h \right)
\right\} \cr
&{\longrightarrow \atop h \to  e^\varphi \eta}
\int [\Scd \L]_\eta \,
\exp \left\{\frac{i}{2} \int d^2 x   \left(\L +
\frac{\varphi}{\sqrt{48 \pi}} \right) \Square_h
\left(\L + \frac{\varphi}{\sqrt{48 \pi}}
\right)  + i S_\ssl[\eta_\mn,\varphi] \right\}, \cr}
\label\scalarsix
\eeq
where the last equality applies in a conformally-flat
background, $h_{\mu \nu}  = e^{ \varphi} \eta_{\mu \nu}$.
After shifting $\L$ to absorb $\frac{1}{\sqrt{48 \pi}}
\, {\Square}_h^{-1} \, {\cal{R}}_h$, or
equivalently $\frac{1}{\sqrt{48 \pi}} \; \varphi$,
we recover in eq.~\scalarsix\
the original massless scalar theory.

We see that although this example has the advantage of
extreme simplicity, it has the drawback that spacetime duality does not do
anything particularly interesting, simply
mapping the massless scalar field theory back onto itself.
This drawback is not shared by the next example, to which we
now turn.

\section{Spacetime Duality II: Bosonization of Fermions}

As our next example we apply spacetime duality to a
free massless fermion in 1+1 dimensions. Since the dual
variable is bosonic, this transformation cannot lead
to the same theory as the one with which we start.

We begin with a massless Dirac spinor $\chi$ in the
presence of a curved background zweibein $e_\mu^a$
which is related to the background metric
in the usual way,
$h_{\mu \nu } = e_\mu^a e_\nu^b \;\eta_{a b}$.
\eq
\eqalign{
&Z[e_\mu^a] = \int [\Scd {\overline{\chi}}]_e
[\Scd {\chi}]_e \; \exp \left\{ i S[e_\mu^a, \ol\chi, \chi] \right\}, \cr
\hbox{where} \qquad &S[e_\mu^a, \ol\chi, \chi]
=  -  \int d^2x  e \; i
\overline{\chi} \, \gamma^\mu
    D_{\mu} \chi   ,\cr}
\label\diracone
\eeq
and $e = \det[e_\mu^a] = \sqrt{- h}$, $\gamma^\mu = \gamma^a
e^\mu_a$, and $D_\mu$ denotes the covariant derivative acting
on spinors.

As for the scalar case, we start the duality program by
introducing a dynamical gravitational field represented
by the zweibein $f_\mu^a = \left( e^{\phi/2}
e_\mu^b \theta_b^a  \right)^\xi$, where $\theta_b^a$ is a local Lorentz
transformation (LLT) and (as before) $\xi^\mu$ parametrizes
diffeomorphisms. The dynamic metric is $g_{\mu \nu } =
f_\mu^a f_\nu^b \; \eta_{a b}$. We are led to the following
extended, gauge-invariant, version of eq.~\diracone
\eq
Z[e_\mu^a] = \int [\Scd {\overline{\chi}}]_f \,
[\Scd {\chi}]_f \, [\Scd  f_\mu^a]_f \;
\Delta_\LM[f_\mu^a,e_\mu^a]
\exp \left\{ i S[f_\mu^a, \overline{\chi}, \chi] \right\} ,
\label\diractwo
\eeq
where the covariant measure for the zweibein integration
is given by
\eq
[\Scd  f_\mu^a]_f  = [\Scd  \phi]_f \,  [\Scd   \xi^\mu]_f
\, [\Scd  \theta^a_b]_f \, \Delta[{\rm gauge \;  fix \; GCT}]
\Delta[{\rm gauge\; fix \;LLT}]
[J_\FP]_f
\eeq

Choosing conformal gauge,  \ie\ ${\xi^\mu} = x^\mu$ and $\theta^a_b
= \delta^a_b$, ensures that ${f_\mu^a} = {e^{\phi/2}  \,
{e_\mu^a} }$. The path integral eq.~\diractwo\ becomes
\eq
\eqalign{
Z[e_\mu^a] &= \int [\Scd {\overline{\chi}}]_{e^{\phi/2} \, e }
\, [\Scd {\chi}]_{e^{\phi/2} \, e} [\Scd
\phi]_{e^{\phi/2} e } \;  [J_\FP]_{e^{\phi/2}\, e } \; \times\cr
&\qquad \qquad \qquad \Delta_\LM[{e^{\phi/2}
\, {e_\mu^a} },e_\mu^a] \exp \left\{ iS[{e^{\phi/2}
\,  {e_\mu^a} },  \overline{\chi}, \chi] \right\} . \cr}
\label\diracztwo
\eeq
The \FPD\ determinant, $J_\FP$, now ensures gauge invariance
with respect  to both GCTs and LLTs, but since LLT gauge
fixing is algebraic and the corresponding Jacobian a
determinant of an orthogonal matrix, $J_\FP$ is the
same as for the scalar theory of \S3.

As usual, the $\phi$ dependence of all measures may be obtained
purely by keeping track of the conformal anomaly, which
for a Dirac fermion is given by
\eq
[\Scd {\overline{\chi}}]_{e^{\phi/2} e }
= [\Scd {\overline{\chi}}]_e \;
e^{\frac{i}{2}  S_\ssl[{h_{\mu\nu}},\phi]}  \qquad
[\Scd {{\chi}}]_{e^{\phi/2} e } = [\Scd {{\chi}}]_e \;
e^{ \frac{i}{2} S_\ssl[{h_{\mu\nu}},\phi]} .
\eeq
Using this directly in eq.~\diracone\ and specializing
 to a conformally-flat
metric $e^a_\mu = e^{\varphi/2} \, \delta^a_\mu$
gives the standard relation between the determinant
of the Dirac operator and the determinant of the
scalar Laplacian
\eq
{\det(i \Dsl_e) \over \det (i \Dsl)}
= {Z[e^a_\mu] \over Z[\delta^a_\mu]}
= {Z\left[ e^{\varphi/2}\, \delta^a_\mu \right]
\over Z[\delta^a_\mu]}
= e^{i S_\ssl[\eta_\mn,\varphi]}
= {[\det (- \Square_h)]^{-1/2} \over [\det (- \Square)]^{-1/2} }.
\label\diracbox
\eeq

We once again choose either eq.~\deltaLMcov\ or
\deltaLMfva\ to define $\Delta_\LM[{e^{\phi/2} {e_\mu^a} },
e_\mu^a] = \Delta_\LM[{e^{\phi} h_\mn }, h_\mn]$.
After substitution of eq.~\deltaLMfva\ into eq.~\diracztwo, the
generating functional becomes
\eq
\eqalign{
Z[e_\mu^a] &= [\det(-\Square_h)] \int [\Scd {\overline{\chi}}]_{e^{\varphi/2}e}
\, [\Scd {\chi}]_{e^{\varphi/2}e} \, [\Scd  \phi]_{e^{\varphi/2}e} \,
[\Scd   \Lambda]_h \quad \times \cr
& \qquad\qquad
\exp \left\{ iS[{ {e_\mu^a} }, \overline{\chi}, \chi] + i
S_\ssl [{ h_{\mu\nu}},\phi]  - i \int d^2 x \sqrt{-h}  \;
\phi \, \Square_h \L\right\} .\cr}
\eeq

Performing the $\L$ and $\phi$ integrations, we recover
the starting point, eq.~\diracone. Instead, evaluating the
path integrals over ${\overline{\chi}},
{\chi}$, and $\phi$, the dual theory becomes
\eq
\eqalign{
Z[e_\mu^a]  &=  \det(i \Dsl_e )
[\det(-\Square_h)]^{1/2} \int [\Scd {\Lambda}]_h \quad
\times \cr
& \qquad \qquad
 \exp  \left\{ \frac{i}{2} \int d^2x \sqrt{ - h}
\, \left(  {\Lambda}  + \frac{1}{\sqrt{48 \pi}}
\, \frac{1}{ {\Square}_h}  \, {\cal{R}}_h \right)
{\Square}_h \left(  {\Lambda}  + \frac{1}{\sqrt{48 \pi}}
\,  \frac{1}{ {\Square}_h}  \, {\cal{R}}_h \right)
\right\} . \cr}
\label\dualfermion
\eeq
Inspection of eq.~\diracbox\ shows that the
$e^a_\mu$ dependence of the functional determinants in
front of this expression cancels.

Eq.~\dualfermion\ is the image of the free Dirac fermion
under spacetime duality.  After a shift, we obtain the path
integral for a massless scalar field.  This may be regarded as a
way of expressing bosonization as a duality transform,
which is an alternative to that of ref.~\bosorefs.

\section{Superduality I: The $(1,1)$ Supersymmetric Wess-Zumino Model}

We now turn to examples where the spacetime symmetry on
which the duality is based is supersymmetry. This introduces
the novel feature that some of the dual variables may now
be fermions. We present two examples of supersymmetry-based
duality, or superduality. In this section we construct
the superdual of the $(1,1)$ supersymmetric Wess-Zumino (WZ)
model with the goal of exhibiting the method in
the simplest possible setting. We find a result similar
to what was found in \S3 above: the dual model is
the same as the starting theory. For this reason we
present a more interesting
second example, the $(2,2)$ supersymmetric
WZ model, in the next section.

The general procedure for superduality once more follows the paradigm
of \S2. Starting with a globally-supersymmetric
model, we gauge the global supersymmetry by coupling the model
to supergravity, and then enforce a gauge-fixing
condition which eliminates the resulting dynamical supergravity
(SUGRA) degrees of freedom. We implement this constraint
using an entire supermultiplet of Lagrange-multiplier
fields. To reach the dual formulation, we
integrate the original matter fields and the SUGRA multiplet
to leave the path integral in terms of the Lagrange multipliers.

\ref\lindstrometal{U. Lindstr\"{o}m, N.K. Nielsen, M. Ro\v{c}ek, and
P. van Nieuwenhuizen,  \prd{37}{88}{3588}. }

We start with the action of the (1,1)-supersymmetric WZ
 model in flat Minkowski space, described by a scalar multiplet
($A$, $\chi$, $N$) with action
\eq
S_0 = \frac{1}{2} \int d^2x \left(  A \Square A  - i
\overline{\chi} \gamma^\rho \partial_\rho \chi +
N^2 \right)  \label\freesusy
\eeq
where $A$ and $N$ are real scalar fields and $\chi$ is
a Majorana fermion.

In what follows it will be  convenient
to introduce a compact matrix notation, as in ref.~\lindstrometal,
which is inspired by the superspace formulation of supersymmetry.
We therefore define
\eq
\rho := \left (\matrix{
A\cr      N\cr        \chi\cr}\right )
\hskip 1cm \bf{T} := \pmatrix{
0 & 0 & 1 \cr
0 & 1 & 0 \cr
{\bf C} & 0 & 0 \cr  }
\hskip 1cm \Theta_0 := \pmatrix{
0 & 0 & - i \gamma^\rho \,\partial_\rho \cr
0 & 1 & 0 \cr
\Square & 0 & 0 \cr   }
\label\matter
\eeq
where $\bf{C}$ is the charge conjugation matrix $- \gamma^0$.
With these definitions the action \freesusy\ can
be rewritten in the compact form
\eq
S_0 =  \frac{1}{2} \int d^2x \; \bar{\rho} \; \Theta_0 \rho ,
\label\compact
\eeq
where the conjugate $\bar{\rho}$ is defined  as $\rho^t \bf{T}$,
the superscript `$t$' denoting transposition.

We next couple to a background supergravity multiplet
$B = \{e_{\mu}^a,\psi_{\mu}^\ssb, S^\ssb\}$, in order
to have arguments to follow after performing the path integral
over the WZ multiplet. The resulting action takes the form
\eq
S(\rho, B)   = \frac{1}{2} \int d^2x  e  \left\{   A \Square_e A  - i
\overline{\chi} D_\rho \gamma^\rho \chi + N^2 +
i \kappa  \overline{\chi} \gamma^\mu \gamma^\nu \psi_\mu^B \partial_\nu A -
\frac{\kappa^2}{8} \; \overline{\chi} \chi \;
\overline{\psi}_\nu^B \gamma^\mu
\gamma^\nu \psi_\mu^B  \right\} ,
\label\sugracomp
\eeq
where $\kappa$ is the supergravity coupling constant.
This last expression can also be rewritten, using matrix notation,
\eq
S  =  \frac{1}{2} \int d^2x \; \bar{\rho} \Theta  \rho ,
\label\matrixnot
\eeq
where now
\eq
\Theta := |e| \pmatrix{
\frac{i \kappa}{2}  \gamma^{\mu}   \gamma^{\nu} \psi_\mu^\ssb \partial_\nu
& 0 &
-i \gamma^\rho \,D_\rho  -\frac{\kappa^2}{8}
\bar{\psi}_\mu^\ssb \gamma^{\nu}   \gamma^{\mu}\psi_\nu^\ssb \cr
  0 & 1 & 0 \cr
 \Square_e & 0 & -\frac{i \kappa}{2 |e|}
\partial_{\mu} |e| \bar{\psi}_{\nu}^\ssb
\gamma^{\mu}\gamma^{\nu}  \cr } .
\label\sugraop
\eeq
 (In our notation,
derivatives in $\Theta$ act on everything which stands to
their right.)
The quantum system of interest is given by the following
path integral:
\eq
Z(B) \equiv  Z[e_\mu^a, \psi_\mu^\ssb, S^\ssb]  = \int [\Scd \rho]_B
\exp \left\{ iS (\rho, B)\right\}
\label\qtdef
\eeq
where the measure is required to be locally supersymmetric
(see \lindstrometal\ for one possible construction).

In superconformal gauge the supergravity multiplet has the form
\foot\twotimes{Notice we deliberately choose here, for later
convenience,
a conformal factor for the zweibein which is twice
what we used in previous sections.}
\eq
e_\mu^a \rightarrow e^{\phi^\ssb} \delta_\mu^a \hskip 1cm \psi_\mu^\ssb
\rightarrow \hf \; \gamma_\mu \psi^\ssb
 \hskip 1cm  S^\ssb \rightarrow S^\ssb ,
\eeq
Because the superconformal gauge supergravity multiplet
 has the same field content as does a scalar
matter multiplet, it is convenient to group
these fields into a
background-field multiplet ${\cal B} = (\phi^\ssb, \psi^\ssb, S^\ssb)$.
For later use we record the superconformal gauge limit
of the operator of eq.~\sugraop:
\eq\label\conflim
\Theta_\SC^{\cal B} = \pmatrix{
0 & 0 & - ie^{\phi^\ssb} \gamma^\rho \, \partial_\rho \cr
0 & e^{ 2 \phi^\ssb} & 0 \cr
\Square & 0 & 0 \cr
}
\eeq
Although equation \conflim\ seems to imply
that the path integral has lost its
dependence on $\psi^\ssb$ and $S^\ssb$, this is
not true once the SUGRA-dependence of the
path-integral measures is taken into account.

 To proceed with the duality construction, we start now with the
scalar multiplet $\rho$ coupled to a dynamical supergravity
multiplet ${T} = (f_\mu^a, \psi_\mu^\sst, S^\sst )$, and a generating
functional
defined by the path integral
\eq
\eqalign{
Z [B] & = \int [\Scd \rho]_\sst [\Scd T]_\sst
\Delta_{LM}[T, B] \exp \left\{ iS [\rho, T]\right\} }
\eeq
where $\Delta_{LM}$ is the constraint which the duality
procedure introduces to trivialize the integral over
the dynamical SUGRA multiplet (\ie\ by setting
 $T=B$).
As before we use the latitude
in choosing this constraint to ensure that all of
the SUGRA dependence of the path-integral measures
is precisely compensated, ensuring that the original
theory is recovered.
 To describe the $T$ integration in detail, we
parametrize the dynamical supergravity fields in terms of
diffeomorphisms, Lorentz transformations, and local supersymmetry
transformations acting on the background fields $B$.  The
integration symbol $[\Scd T]_\sst$ stands for integration over
the corresponding transformation  parameters, as well as
 a gauge-fixing condition  and the
 Fadeev-Popov determinant ${\cal J}_{FP}$.  This is done most
easily, as in the ordinary gravitational case, by {\it choosing}
superconformal gauge for the background fields and {\it imposing}
superconformal gauge for the dynamical fields, in which case
one has essentially a linear splitting  $f_\mu^a = e^{{\phi}^\ssq}
e_\mu^a = e^{{\phi^\ssq} + {\phi^\ssb}}\delta_\mu^a$,
 $\psi_\mu^\sst = f_\mu^a \gamma_a
(\psi^\ssb + \psi^\ssq)$, $S^\sst = S^\ssb +S^\ssq$.
\foot\elaboration{This point requires some elaboration:
 in a general gauge, starting with the dynamical supergravity multiplet
one would introduce the background supergravity by a standard,
 but nonlinear,
background-quantum splitting.  In superconformal gauge however, where
the superspace description is by means of a  scalar compensator
superfield ,  the splitting is linear.}
We group the superconformal gauge dynamical
SUGRA fields into another scalar  multiplet ${\cal Q} = (\phi^\ssq,
\psi^\ssq, S^\ssq)$. With this choice the total SUGRA multiplet enters
into the WZ action only as the sum of a background and a dynamical
contribution ${\cal B} + {\cal Q}$.

We choose the Lagrange multiplier fields as members of
another scalar multiplet $\Lambda = (L
,\eta, F)$. We are led to the
following representation of eq.~\qtdef:
\eq
Z [{\cal B}]  = \int [\Scd \rho]_{{\cal B} + {\cal Q}}
[\Scd {\cal Q}]_{{\cal B} + {\cal Q}}  [\Scd \Lambda]_{{\cal B} + {\cal Q}}
\sdet[ \Theta_{SC}^{{\cal B} + {\cal Q}} ]
\exp \left\{ iS [\rho;{{\cal B} + {\cal Q}} ] - i \int
d^2 x \bar{{\cal Q}} \Theta_{SC}^{{\cal B}} \Lambda \right\}
\label\qtggd
\eeq
where the superdeterminant is the remnant of
$\Scj_\FP \;
\Delta_\LM$.\foot\relation{
In superspace we have the  relation for the spinor vielbein
 $\ss E_\alpha ({ total} )=
e^{\cal Q} E_\alpha ({ background})$ and  $\ss {E_T}^{-1}R_T -{E_B}^{-1}R_B
= -4{E_B}^{-1} {\nabla}_B^2 {\cal Q}$. The Lagrange multiplier
action $\ss \int d^2x d^2\theta {E_B}^{-1} \Lambda {{\nabla}_B}^2 {\cal Q}$
reduces at the component level  to the term in eq. \qtggd.}
  Using superconformal invariance
one can replace $\Theta_{SC}^{{\cal B}}$ with
$\Theta_\SC^{{\cal B} + {\cal Q}}$ in the Lagrange multiplier term
for ease of performing the duality algorithm.

To verify that eqs.~\qtggd\ and \qtdef\ are equivalent, we
perform explicitly the $[\Scd \Lambda]$ integral, obtaining
the functional delta function
$\Delta[\Theta_\SC^{{\cal B} + {\cal Q}} {\cal Q}] =
\sdet[\Theta_\SC^{{\cal B} + {\cal Q}} ]^{-1} \Delta[ {\cal Q}]$.
Eq.~\qtdef\ is then obtained by using this delta
function to integrate out $ {\cal Q}$.

\ref\fradkintseytlin{E. Fradkin and A. Tseytlin, Phys. Lett.
106 B (1981) 63.}

\ref\slaction{
A. M. Polyakov,  \plb{103}{81}{211};  A. M. Polyakov, {\it Gauge Fields and
Strings}, Harwood, 1987; see also refs. \fradkintseytlin and \lindstrometal.
 }

To reach the dual formulation, we integrate all fields
except the Lagrange multiplier multiplet.
The matter multiplet integrates to give $\sdet
[\Theta_\SC^{{\cal B} + {\cal Q}}]^{-1/2} $
which changes the power of the superdeterminant term
in the path integral. In order to perform the
$\Scq$ integration, we need to make explicit the dependence
of the measures on this variable. As in the
scalar case, we do so by performing a super-Weyl transformation
and rewriting the superdeterminant as in eq.~\xresult
\eq
[\Scd {\cal Q}]_{{\cal B} + {\cal Q}} \; [\Scd \Lambda]_{{\cal B} + {\cal Q}}
\;
 \sdet[ \Theta_\SC^{{\cal B} + {\cal Q}} ]^{1/2}
= [\Scd  {\cal Q}]_{0} \; [\Scd \Lambda]_{0} \;
\sdet[ \Theta_{0} ]^{1/2} \; \exp\Bigl[ i S_\SL({{\cal B} + {\cal Q}}) \Bigr] ,
\label\wrescale
\eeq
where the subscript `0' on the measures indicates that all of the
dependence on the background SUGRA fields has been
made explicit, leaving measures which depend only
on the flat metric.
Here $S_\SL({{\cal B} + {\cal Q}})$ denotes the super-Liouville
action. For the superconformally-flat SUGRA configurations
we are considering  (keeping in mind our unconventionally
normalized conformal factor, $g_\mn = e^{2\phi} \; \eta_\mn$)
the super-Liouville action is given by \slaction:
\eq
\eqalign{
S_\SL({\cal B}) &= - \; \frac{1}{8 \pi} \int d^2 x \; \left( \frac12
\phi \,\Square \,\phi  -\;  \frac{i}{2} \bar \psi
\gamma^\mu \partial_\mu \psi + \frac{1}{2}\, S^2 \right) \cr
&= - \; \frac{1}{16\pi} \int d^2 x \; \bar{{\cal B}} \Theta_0 \, {\cal B}  .
\cr}
\label\SLaction
\eeq

These steps lead to the following expression
\eq
\eqalign{
Z [{\cal B}]  &= \int [\Scd {\cal Q}]_{0} [\Scd \Lambda]_{0}\;
\sdet[ \Theta_0]^{1/2}
\exp \left\{  - i\int d^2 x \; \bar{{\cal Q}} \Theta_0
\Lambda  - \;  \frac{i}{16 \pi}
\int d^2 x \;{(\bar{\cal B} + \bar{\cal Q})} \Theta_0
({\cal B} + {\cal Q} ) \right\}  \cr
&= \int [\Scd {\cal Q}]_{0} [\Scd \Lambda]_{0} \;
\sdet[ \Theta_0]^{1/2}
\exp \left\{ -\;   \frac{i}{16\pi} \int d^2 x  \;
\left(\bar{\cal Q}+ \bar{\cal B}  + 8 \pi \, \bar{\Lambda} \right)
\Theta_0 \left( {\cal Q} + {\cal B}  + 8 \pi\,  \Lambda
\right)  \right.  \cr
&  \qquad\qquad\qquad\left. + \frac{i}{16\pi} \int d^2 x \;
\left( \bar{\cal B}  + 8 \pi \, \bar\Lambda \right) \Theta_0 \left(
{\cal B} + 8 \pi \, \Lambda \right)   -\frac{i}{16\pi } \int d^2 x \;
 \bar{{\cal B}}  \Theta_0 {\cal B}   \right\} , }
\label\sugraeff
\eeq
where the last line is obtained by completing the squares.
Performing the ${\cal Q}$ path integral removes the ${\cal Q}$-dependent
first term from the action, and produces a superdeterminant which
cancels the superdeterminant which is already present.
Recognizing the last term in the action as the super-Liouville action,
we may absorb it to rewrite the $\Lambda$ measure in terms
of the SUGRA background, ${\cal B}$. The remaining
term is the dual action, which takes a canonical form after
appropriately shifting and rescaling the multiplet $\Lambda$.
We are left with the final dual form
\eq
Z [{\cal B}]  =   \int [\Scd \Lambda]_{{\cal B}} \;
\exp \left\{   \frac{i}{2} \int d^2 x \; \bar{\Lambda}
\Theta_\SC^{{\cal B}} \Lambda \right\} .
\eeq

Our result is similar to the scalar example of \S3;
the dual action is identical to the original theory with which we started.

\section{Superduality II: The $(2,2)$ Supersymmetric Wess-Zumino Model}

For our final example we choose the simplest model
for which the superduality transformation may be explicitly
performed, and yet for which the result differs nontrivially
from the original model. We consider the $(2,2)$
generalization of the previous example -- a single massless WZ
multiplet coupled only to background SUGRA fields.
\ref\twisted{ S.J. Gates,
C.M. Hull and M. Ro\v{c}ek,  \npb{248}{84}{157}.}
\ref\twotworefs{
See ref. \rocekverlinde and E. Witten, \npb{403}{93}{159}.
}
\ref\marciamarc{M.T. Grisaru and M.E. Wehlau, \npb
457 (1995) 219.}
\ref\marciamarcjim{S.J. Gates Jr., M.T. Grisaru and M.E. Wehlau,
\npb{460}{96}{579}.}
The feature which makes the (2,2) example more
interesting  is the fact that (2,2) scalar
multiplets come in more than one type  \twisted, \twotworefs,
\marciamarcjim.  We shall find that superduality
can map one type of multiplet into  another.

The basic matter multiplet for (2,2)
supersymmetry has field content $(\phi, \psi, F )$,
where $\phi$ and $F$ are complex scalars and $\psi$
is a Dirac spinor. In (2,2) superspace these fields may be
grouped into a scalar superfield $\Phi$ satisfying
 a supersymmetric constraint.
Two types of constraints are relevant for us.
 For global (2,2) supersymmetry
a {\it chiral} scalar superfield is one which
satisfies $D_{\stack{.}{+}}
\Phi = D_{\stack{.}{-}} \Phi = 0$. Here $D_\pm$
are the supercovariant spinor derivatives, whose
complex conjugates are denoted $D_{\dot{\pm}}$.
A {\it twisted-chiral} scalar superfield is defined
by $D_{\stack{.}{+}} \Phi = D_{{-}} \Phi = 0$.
Similar definitions apply for local (2,2) supersymmetry,
with the derivatives $D_\alpha$ and $D_{\dot\alpha}$
replaced by their locally supercovariant counterparts,
$\nabla_\alpha$ and $\nabla_{\dot\alpha}$.

To apply the duality algorithm we require the coupling
of these multiplets to \twtw\ supergravity. The irreducible  \twtw\
supergravity multiplet has the field content $(
e_\mu^a, \xi_\mu, A_\mu, G)$, where $e_\mu^a$ is the zweibein,
$\xi_\mu$ is a Dirac gravitino,
$G$ is a complex scalar auxiliary field and $A_\mu$ is a gauge potential.
$A_\mu$ gauges one of the two $U(1)$ internal symmetries of
 the \twtw\ supersymmetry algebra.
\ref\howepap{ P.S. Howe and G. Papadopoulos ,  {\it Class. Quantum Grav.}
 {\bf 4} (1981) {11}.}
\ref\gatesliu{S.J. Gates Jr.,  L. Liu and R. Oerter, \plb
{218} {89} {33}.}
These symmetries
act as vector and  axial  symmetries  on the various
fermions within \twtw\ supermultiplets. There are
two distinct \twtw\ supergravity multiplets, denoted by
 $U_\ssv(1)$ or $U_\ssa(1)$ supergravity depending on which one
 of these symmetries is gauged by the field $A_\mu$ \howepap, \gatesliu.
For both of
these supergravities the Ricci scalar $\Scr(e_\mu^a)$
lies
within a scalar superfield $R$. For $U_\ssa(1)$ supergravity
 this superfield is
chiral, while for $U_\ssv(1)$ supergravity it is twisted-chiral.

In principle there are four \twtw ~ WZ models
to consider, depending on whether the matter multiplet
is chiral ($C$) or \twch ($T$), and whether we use
$U_\ssa(1)$ or $U_\ssv(1)$ supergravity.  We denote
these four possibilities by $CA$, $CV$, $TA$ and
$TV$.
\ref\brinkschwarz{L. Brink and J. H. Schwarz, \npb{121}{77}{285}.}
The lagrangian for each of these four possibilities is known,
both in superspace  \marciamarc, \marciamarcjim\ and in components
\fradkintseytlin, \brinkschwarz.
In superspace the action distinguishes between the four cases, whereas
the component action (in the absence of masses and self-interactions)
does not. It is given by
\eq
\eqalign{
S =& \frac{1}{2} \int d^2 x | e | \left\{
-g^{\mu \nu} \partial_\mu \phi \partial_\nu \phi^* - {i} \bar{\psi}
\gamma^\mu D_\mu \psi -2 A_\mu \bar{\psi} \gamma^\mu \psi -2 \left(
\partial_\mu \phi^* + \bar{\psi} \xi_\mu \right) \bar{\xi}_\rho \gamma^\mu
\gamma ^\rho \psi  + \right. \cr
& \qquad \left. -  2 \left( \partial_\mu \phi + \bar{\xi}_\mu \psi \right)
\bar{\psi} \gamma ^\rho \gamma^\mu \xi_\rho  + F^* F \right\}
\label\twotwoaction }
\eeq

\ref\moretocome{M. T. Grisaru and M. E. Knutt-Wehlau, in preparation.}

The rest of this section applies superduality to the four
versions of the massless model with no self-couplings.
We show that superduality interchanges these
multiplets in the following way
\eq
\eqalign{
CA \to CA \qquad&\qquad TA \to CA \cr
CV \to TV \qquad&\qquad TV \to TV .\cr}
\label\sdualmaps
\eeq
Because the  component actions
do not distinguish between
the various kinds of multiplets,
we examine the consequences of superduality in a
superspace formulation. There are, however, some subtleties
concerning  the path-integral
measures for \twtw\ superspace \moretocome,  and consequently we
shall proceed  in two
steps. In order to show that the duality procedure produces
a well-defined and local result --- which need not be
generally true --- we first perform the duality algorithm on
 the component action, keeping explicit track of the cancellation of
all nonlocal functional determinants.
We follow this component calculation by
presenting the duality argument directly in superspace yielding
the results as indicated in
eq.~\sdualmaps.

\subsection{$(2,2)$ Superduality in Components}

 As in previous sections,
we start with matter coupled to background SUGRA
\label\susycompone
\eq
Z(B) = \int \left[
\Scd(M,{\bar M})\right]_B \exp \left\{
+ i \int d^2 x {\bar M} \Theta^B M \right\} ,
\eeq
where the action is taken from eq.~\twotwoaction, but is
written in a compact notation which is similar to
that used in the (1,1) case. $M$ collectively
denotes the matter multiplet, $(\phi, \psi, F)$,
and a bar on $M$ indicates hermitian
conjugation and multiplication by a generalization of the
charge conjugation matrix ${\bf T}$. The operator $\Theta^B$
depends on the background supergravity multiplet $(
e_\mu^a, \xi_\mu, A_\mu, G)$.
\ref\spinstring{M. Roc\v{e}k, P. van Nieuwenhuizen,
and S.C. Zhang, \anp{172}{86}{348}.}
 In superconformal gauge \foot\gauge{The
terminology is not strictly correct.  In superspace,
superconformal gauge is defined by setting the prepotential
 $\ss H^a$ to zero  and  keeping the conformal compensator superfield.
  Here one first goes to WZ gauge by setting the conformal
compensator to 1 and gauging away some components of $\ss H^a$, and then
going to component conformal gauge.
  See also the discussion in  appendix B of \spinstring~. }
 this multiplet reduces to the conformal factor  $\sigma$ of the metric,
a Dirac spinor $\lambda$ from the gravitino,
 $\xi_\mu = \hf \gamma_\mu \lambda$,
the transverse component of the gauge field,
$A_\mu = \hf \; \epsilon_\mn \, \partial^\nu \rho$,
 and
the auxiliary field $G$, which we collectively denote by
${\cal B} = (\sigma, \lambda, \rho, G)$.
The fields $\sigma$, $\rho$ can be combined into a single complex
scalar field.
 Notice that the SUGRA
field content in this gauge is the same
as for the matter multiplet.

We imagine the ${\cal B}$-dependence of
the measure in eq.~\susycompone\ to be defined to
ensure invariance with respect to local \twtw\ supersymmetry.
We can infer this dependence from the superconformal anomaly
of this model \fradkintseytlin:
$\left[ \Scd(M, {\bar M}) \right]_{{\cal B}} =
\left[\Scd (M, {\bar M}) \right]_0 \;
\exp\Bigl\{ i S_\SL({\cal B}) \Bigr\}$,
 where~\foot\appendixfoot{See appendix for a discussion of the sign of
the second term.}
\eq
\eqalign{
S_\SL({\cal B}, \bar{\cal B}) &= - \; \frac{1}{4 \pi}
 \int d^2 x \; \left( -\; \frac12
\partial_a \sigma \,\partial^a \sigma - \; \frac12
\partial_a \rho \,\partial^a \rho -\; \frac{i}{2} \bar \lambda
\gamma^a \partial_a \lambda + \frac{1}{2} G^*
G \right) \cr
&= - \; \frac{1}{4\pi} \int d^2 x \; {\bar{\cal B}} \,
\Theta_0 \,{\cal B} , \cr}
\label\anomaction
\eeq
which is once again simply proportional to the kinetic action for a
 matter multiplet.

Superduality proceeds by introducing dynamical supergravity
fields. After imposing  superconformal gauge we are left with
the matter multiplet coupled to
a total SUGRA multiplet
${\cal B} +{\cal  Q}$. We must also constrain away the ${\cal  Q}$
degrees of freedom,
 to ensure consistency with the original form,
eq.~\susycompone. We do so using a multiplet of Lagrange
multipliers, denoted $\Lambda  = (L_1+i
L_2, \eta, {\cal G} )$. Note that $\L$ has the
same field content as have the superconformal
supergravity multiplets, ${\cal B}, {\cal Q}$, and the matter multiplet, $M$.
We are led in this way to the expression:
\eq
\eqalign{
Z({\cal B}, {\bar{\cal B}}) &= \int \left[
\Scd (M,{\bar M}) \right]_{{\cal B}+{\cal Q}} \left[
\Scd ({\cal Q}, {\bar{\cal Q}})\right]_{{\cal B}+{\cal Q}}
\left[ \Scd(\Lambda,{\bar{\Lambda}})\right]_{{\cal B}+{\cal Q}}
(\sdet[\Theta^{{\cal B}+{\cal Q}}_{SC}])^2 \cr
& \qquad \times \exp \left\{
+ i \int d^2 x {\bar M} \Theta^{{\cal B}+{\cal Q}}_{SC} M + i \int d^2 x
\left(
{\bar{\cal Q}} \Theta^{\cal B}_{SC} \Lambda + {\bar{\Lambda}}
\Theta^{\cal B}_{SC}{\cal Q} \right) \right\} . \cr }
\label\susycompnewone
\eeq
Here the subscript on the field operator $\Theta_{SC}$ is
a reminder that it is obtained from eq.~\twotwoaction\ in
superconformal gauge.  Again, because of superconformal invariance
of the action, one can replace $\Theta^{\cal B}_{SC}$ by $\Theta^
{{\cal B}+{\cal Q}}_{SC}$ in the Lagrange multiplier term.

To perform the relevant path integrals, one can
make all SUGRA dependence explicit by performing a
superconformal transformation. When this is done for
eq.\ \susycompnewone, the transformation of the measures
for ${\cal Q}$ and $\Lambda$ cancels a factor from  the
superdeterminant, leaving only a single factor of the
anomaly action coming from $[\Scd (M, {\bar M})]$. We find
\eq
\eqalign{
Z({\cal B}, {\bar{\cal B}}) = &\int \left[\Scd (M,{\bar M})\right]_0
\left[\Scd ( {\cal Q}, {\bar{\cal Q}})\right]_{0}
\left[\Scd (\Lambda, {\bar{\Lambda}})\right]_{0}
(\sdet[\Theta_0])^2 \times \cr
&
\exp \left\{
+ i \int d^2 x {\bar M} \Theta_0 M + i \int d^2 x \left(
{\bar{\cal Q}}  \Theta_0 \Lambda +
{\bar{\Lambda}} \Theta_0 {\cal Q} \right)+ i S_\SL ({\cal B} +{\cal Q})
\right\} }
\label\susycomptwo
\eeq

We first verify the equivalence of eq.~\susycomptwo\ with
our starting expression, eq.~\susycompone. To do so, we
change variables $\L \to \Theta_0 \L$, and integrate
 to obtain the functional delta function $\Delta[{\cal Q}]$.
Using this to perform the ${\cal Q}$ integration leaves
\eq
Z({\cal B}, {\bar{\cal B}}) = \int \left[\Scd (M, {\bar M})\right]_0 \;
\exp \left\{
+ i \int d^2 x {\bar M} \Theta_0 M + i S_\SL ({\cal B})
\right\} .
\label\susycompcheck
\eeq
This reproduces eq.~\susycompone\ once $S_{SL}({\cal B})$ is rescaled into
the measure using a superconformal transformation.

To obtain the dual, we evaluate the path integral
over $(M, {\bar M})$ to obtain $\sdet [ \Theta_0 ]^{-1}$
which partially cancels the explicit superdeterminant
 already present in eq.~\susycomptwo. This leaves
\eq
\eqalign{
Z({\cal B}, {\bar{\cal B}}) = &\int \left[\Scd ({\cal Q},
 {\bar{\cal Q}})\right]_{0}
\left[\Scd (\Lambda, {\bar{\Lambda}})\right]_{0}
\sdet[\Theta_0]\times \cr
&
\exp \left\{
 i \int d^2 x \left(
{\bar{\cal Q}}  \Theta_0 \Lambda +
{\bar{\Lambda}} \Theta_0 {\cal Q} \right)- \frac{i}{4 \pi} \int d^2
x ( {\bar{\cal B}} + {\bar{\cal Q}} ) \Theta_0  ({\cal B} + {\cal Q})
\right\} . }
\label\susycompthree
\eeq
Completing the squares in the exponent allows it to be rewritten as
\eq
\eqalign{
i S = - \; \frac{i}{4 \pi} & \int d^2 x \left(
\bar{{\cal Q}} +  \bar{{\cal B}} -   4 \pi \,
\bar{\Lambda} \right) \Theta_0   \left(
  {\cal Q}  +  {\cal B} - 4 \pi \, {\Lambda} \right) \cr
& + \frac{i}{4\pi} \int d^2 x \left(\bar {\cal B}  -  4 \pi \,
\bar{\Lambda} \right) \Theta_0
\left(  {\cal B} -  4 \pi \,  \Lambda
\right) - \frac{i}{4 \pi} \int d^2 x \bar{{\cal B}} \,
\Theta_0 \, {\cal B} }
\eeq
Rescaling $\Lambda \rightarrow  \Lambda/ \sqrt{ 4 \pi}$, and
performing the ${\cal Q}$ path integration, we obtain
\eq
\eqalign{
Z({\cal B}, {\bar{\cal B}}) = &  \int [\Scd (\Lambda, {\bar \Lambda})]_0 \,
\exp \left
\{- \frac{i}{4 \pi} \int d^2 x \bar{{\cal B}}
\Theta_0 {\cal B}  \right\}\, \cr
& \qquad \qquad \times \exp \left\{ i \int d^2 x \left(
\bar{\Lambda} - \frac{\bar \Scb}{\sqrt{4 \pi}}  \right)
\Theta_0  \left( \Lambda - \frac{\Scb}{\sqrt{4 \pi}}
\right) \right\} , \cr}
\eeq
which, after shifting the dual multiplet $\L$ to absorb
${\cal B} /\sqrt{4 \pi}$ and rescaling
the first exponential   back into the measure
  $[\Scd ( \L, {\bar \Lambda})]_0 $,
we recognize as the generating functional for a massless (2,2)
matter multiplet:
\eq
Z({\cal B}, {\bar{\cal B}}) = \int
[\Scd ( \Lambda,{\bar \Lambda})]_{{\cal B}} \exp \left\{ i
\int d^2 x   \; \bar{\Lambda}   \Theta_{SC}^{\cal B}  \Lambda
\right\}
\eeq

Superficially, it appears that we have obtained our starting
action, eq.~\susycompone, although this need not be true because the
  component action
  does not distinguish  between chiral and twisted-chiral
 multiplets.
The distinction might become visible if one were to use more
complicated actions,
such as having more than one matter multiplet, or including self-interactions.
Alternatively, one can examine the situation in
superspace, as we do next.

\subsection{Superduality in Superspace}

To see how  the mappings in eq.~\sdualmaps\ arise we
 examine the \twtw-invariant form of the Lagrange multiplier
action used in \S3, which involved terms
of the form $\sqrt{-g} \, \L \, \Scr_g$. Recall that in superspace
$\Scr_g$ lives in a scalar superfield $R$, which is chiral
in $U_\ssa(1)$ supergravity and twisted-chiral in $U_\ssv(1)$ supergravity.
The \twtw-supersymmetric generalization of  the condition that imposes
$\sqrt{-g}{\cal R}_g - \sqrt{-h} {\cal R}_h=0$
involves a Lagrange-multiplier superfield with the same
chirality properties as $R$ itself, as follows: In $U_\ssa(1)$ supergravity
and superconformal gauge  we write  for the full superspace vielbein
 $E_{\alpha}^T = e^{\cal Q} E_{\alpha}^B$, where the {\it  chiral} superfield
 ${\cal Q}$ plays the analogous role
to $\phi$ in \S3.  Thus, the constraint ${\cal E}_T^{-1} R_T  -
 {\cal E}_B^{-1} R_B =
 {\cal E}_B^{-1} \bar{\nabla}^2_B \bar{\cal Q}=0$
($\Sce$
 is the density superfield which supercovariantizes
the chiral integration)
is enforced by a chiral integral
\eq
\int d^2 x d^2\theta \; \Sce_B^{-1} \; \Lambda \, {\bar{\nabla}}_B^2 {\bar
{\cal Q}} =\int d^2x d^4 \theta \,  E^{-1} \Lambda {\bar {\cal Q}}
\label\sinvlmterm
\eeq
where $\Lambda$ is a chiral superfield and $E^{-1}$ is the (nonchiral)
 superdeterminant of the vielbein for the full superspace
integral.\foot\convs{Our conventions in this
section
follow those of refs.~\marciamarc\ and \marciamarcjim.}
Correspondingly in $U_\ssv(1)$ theory, $R$,  the Lagrange multiplier
 $\Lambda$, and the density are all twisted chiral.
Because the Lagrange multipliers end up being the dual fields,
 superduality takes both $C$ and $T$ multiplets into $C$ multiplets
 in $U_\ssa(1)$ supergravity.
Conversely in $U_\ssv(1)$  supergravity,
superduality always produces a $T$ multiplet as in eq.~\sdualmaps.
We now demonstrate this explicitly.

Consider, for example, a  twisted-chiral superfield ${\cal X}$ in
$U_\ssa(1)$ supergravity. The invariant generating functional
is defined by
\eq
Z = \int \left[
\Scd({\cal X},{\bar{\cal X}})\right]_{E_{A}^{M}} \exp
\left\{
- i \int d^2 x \, d^4 \theta \;
E^{-1} \; {\bar{\cal X}} {\cal X}
\right\}
\label\susytwo
\eeq

Performing the functional integral over  ${\cal X}$ involves some
subtleties that we will not address here (see however  \moretocome).
In $U_A(1)$ supergravity one finds \marciamarcjim:
\eq
\eqalign{
Z &=\left[ \det {\Square}_+ \right]^{-\hf }
 \cr
& =  \exp \left[ \frac{i}{8 \pi} \int d^2 x \, d^4 \theta
\; E^{-1}\; \left({\bar R} \frac{1}{{\Square}_+} R \right)\right] \cr
& \to  \exp \left[- \frac{2i}{\pi} \int d^2 x \, d^4 \theta
\; \bar{\Sigma} {\Sigma} \right],
\qquad\qquad\hbox{(in superconformal gauge)} ,\cr}
\label\sspanomaction
\eeq
Here ${\Square}_+ = \bar{\nabla}^2 \nabla^2$ is the  superspace
d'Alembertian acting on a chiral superfield  \ref\book{ S.J. Gates,
Jr., M.T.Grisaru, M. Ro\v{c}ek, and W. Siegel, {\it Superspace},
(Benjamin-Cummings), 1983.} and, generically in
superconformal gauge,  $R = -4 {\bar{\nabla}}^2 {\bar \Sigma}$,
 ${\bar R} = 4 {\nabla}^2 {\Sigma}$, where $\Sigma$ is a chiral
 (compensator)
superfield (denoted $\sigma$ in refs. \marciamarc, \marciamarcjim).

We now start dualizing.
We   choose the  gauge for  dynamical
supergravity  such that $E_\alpha ({\rm dynamical}) =
e^{\cal Q}E_\alpha ({\rm  background})$ and in addition we put the
 background in conformal gauge  $E_\alpha  ({\rm  background}) =
 e^{\cal B} D_\alpha$ so that again we are dealing with a linear
splitting ${\cal B} +{\cal Q}$ in terms of chiral superfields. As discussed
above, we also introduce
a Lagrange-multiplier chiral superfield $\Lambda$ as in eq.~\sinvlmterm,
We choose the following gauged version of eq.~\susytwo:
\eq
\eqalign{
& Z({\cal B}, {\bar{\cal B}}) = \int \left[
\Scd({\cal X}, {\bar{\cal X}})\right]_{{\cal B}+{\cal Q}} \left[
\Scd({\cal Q}, {\bar{\cal Q}})\right]_{{\cal B}+{\cal Q}}
\left[
\Scd(\Lambda, {\bar{\Lambda}})\right]_{{\cal B}+{\cal Q}}
  \det[ {\Square}_+^{{\cal B}+{\cal Q}}]\cr
& \qquad \times \exp \left\{
- i \int d^2 x \, d^4 \; \theta
 {\bar{\cal X}}  {\cal X} + i
\int d^2 x \, d^4 \theta \;
({\bar{\cal Q}} \Lambda + {\bar{\Lambda}} {\cal Q} )
\right\}
}
\label\susyLM
\eeq
where the determinant was chosen by
the requirement that eq.~\susyLM\ reduce to eq.~\susytwo,
once the $\Lambda$ and ${\cal Q}$ are functionally integrated.
To see this, one rewrites the Lagrange multiplier term as
$i\int d^2x d^2 \theta \; \Sce_B^{-1} \;  {\cal Q} \bar{\nabla}^2
\bar{\Lambda}+
i\int d^2x d^2 \bar{\theta} \;\bar{ \Sce}_B^{-1} \;  \bar{\cal Q} {\nabla}^2
 {\Lambda}$
and makes the change of variables $\Lambda  \to
 \bar{\nabla}^2 \bar{\Lambda}' $, $\bar{\Lambda}  \to
 {\nabla}^2 {\Lambda}' $ with Jacobian $(\det   \bar{\nabla}^2 \times
\det   {\nabla}^2 )^{-1}= (\det \Square_+)^{-1}$. This cancels the
determinant factor already present and produces delta-functions
that set ${\cal Q}=0$, \etc
This guarantees the required
equivalence of eqs.~\susyLM\ and \susytwo.

To reach the dual formulation we instead
integrate ${\cal X}$ and ${\cal Q}$, leaving only $\Lambda$ as
the dual field.  As in eq.~\sspanomaction,
the ${\cal X}$ integral gives a factor of
$( \det[ {\Square}_+^{{\cal B}+{\cal Q}}])^{-\hf}$
and Weyl rescaling the  ${\cal Q}$ and $\Lambda$ measures gives
$\left[\Scd({\cal Q}, {\bar{\cal Q}})\right]_{{\cal B}+{\cal Q}}
\left[
\Scd(\Lambda, {\bar{\Lambda}})\right]_{{\cal B}+{\cal Q}}=
\left[\Scd({\cal Q}, {\bar{\cal Q}})\right]_{0}
\left[
\Scd(\Lambda, {\bar{\Lambda}})\right]_{0}
( \det[ {\Square}_+^{{\cal B}+{\cal Q}}])^{-1}$. We obtain
\eq
\eqalign{
Z({\cal B}, {\bar{\cal B}}) &=\int
 \left[
\Scd({\cal Q}, {\bar{\cal Q}})\right]_{0}
\left[\Scd (\Lambda, {\bar{\Lambda}})\right]_{0}
  \exp \left\{
 i\int d^2 x \, d^4 \theta \;
({\bar{\cal Q}} \Lambda + {\bar{\Lambda}} {\cal Q} )
\right\}\cr
&\times  \exp \left\{
 -\frac{2i}{\pi}\int d^2 x \, d^4 \theta \;
({\bar{\cal Q}}+\bar{\cal B}) ( {\cal Q}  +{\cal B})
\right\}
}
\eeq
where the last exponential is a representation of
$( \det[ {\Square}_+^{{\cal B}+{\cal Q}}])^{-\hf}$, {\it c.f.}
eq.~\sspanomaction.

Suitably completing squares,
 performing the
Gaussian ${\cal Q}$ integral and rescaling $\Lambda$ finally gives:
\eq
Z( {\cal B}, {\bar{\cal B}} ) =  \int   \left[
\Scd(\Lambda, {\bar{\Lambda}})\right]_{({\cal B})}
 \exp \left\{  i \int d^2 x d^4 \theta
  \left(   {\bar{\Lambda}} -
\frac{2 \bar \Scb}{\pi}   \right)  \left(   \Lambda -
\frac{2 \Scb}{\pi} \right) \right\} ,
\eeq
which we recognize, after a shift, as the superspace generating
functional for the
massless chiral multiplet $\Lambda$.

We have transformed the twisted-chiral starting
multiplet, ${\cal X}$, into a chiral one. As is clear from
the derivation, so long as we stick to $U_\ssa(1)$ supergravity,
a chiral dual variable $\Lambda$ would
also have followed if we had  started with chiral ${\Phi}$.
Similar manipulations for $U_\ssv(1)$ supergravity fill out the
rest of the relationships of eq.~\sdualmaps.

\section{Conclusions}

With this last calculation we have completed what we set out to accomplish.
Four simple (1+1)-dimensional
examples have been the vehicles
for showing how duality transformations can be performed
based on spacetime symmetries, rather than on internal symmetries.

For the examples of \S3\ and \S4, the duality transformation
was based on the general covariance of the starting model. In \S3,
application to a free scalar field gave a trivial result: the
dual and starting models are identical. A nontrivial result
is obtained in \S4\ by starting with a free fermion. Here
the dual is a free boson, giving bosonization in yet
another guise.

The examples in \S5\ and \S6\ focus on supersymmetric models, where
supersymmetry
is used as the symmetry on which duality is based. When applied
to the massless \onon-supersymmetric WZ model in \S5, we find
the dual is equivalent to the starting theory. More interesting
consequences arise when superduality is applied to \twtw-invariant models
in \S6. In this case we find that both chiral and twisted-chiral
multiplets dualize to chiral multiplets if the $U_\ssa(1)$
formulation of supergravity is used. Using instead $U_\ssv(1)$
supergravity one finds both chiral and twisted-chiral multiplets
go to a twisted-chiral dual.
Because mirror symmetry for Calabi-Yau manifolds involves the interchange
of chiral and twisted-chiral multiplets, it would be interesting to find a
closer connection of our results with this case, by investigating
superduality for self-interacting $(2,2)$ scalar multiplets..

\vfill
\eject

\bigskip
\centerline{\bf Acknowledgments}
\bigskip
C.P.B., P.P. and F.Q. thank l'Universit\'e
de Neuch\^atel for fostering the pleasantly
stimulating atmosphere in which this work was started.
We acknowledge with thanks the financial support
our research receives from N.S.E.R.C.\ of Canada,
DGAPA-UNAM Proyect Num. IN103997 and
the UNAM/McGill collaboration agreement. M.E.K was supported in part
by an N.S.E.R.C. postdoctoral fellowship. The research
of M.T.G. is supported in part by NSF Grant No.  PHY-960457.

\appendix{A}{Signs in the super-Liouville Action}

\ref\bg{
See ref. \fradkintseytlin\ and
H. Leutwyler, \plb{153}{85}{65}.}
We discuss here a sign mismatch between the component $(2,2)$
Liouville action that we have used in this work, and the one in
refs.~\bg.
The difference lies in the
relative sign between, for example, the $\partial_\mu \rho \partial^\mu \rho$
and $\partial_\mu \sigma \partial^\mu \sigma$ term in eq. \anomaction. In our
action all signs are the same,  as also follows from superspace
considerations, giving an anomaly multiplet which
is proportional to (but opposite in overall sign from)
the kinetic action of a matter multiplet. By contrast,
 the results in refs. ~\bg\
assign the opposite sign to the $\partial_\mu \rho \partial^\mu
\rho$ term of the anomaly action.

\ref\gaugeanom{J. Schwinger, {\it Phys. Rev.} {\bf 128} (1962) 2425;\bk
For a review with references see: R. Jackiw, in {\it Relativity, Groups and
Topology II}, ed. by B. DeWitt and R. Stora, (North Holland, Amsterdam),
1983.}

We believe this discrepancy  has to do with the interpretation
of the anomaly action, which was used in the euclidean-signature
calculations of refs.~\bg. To see what is going on we note that the
$\rho$-dependent term really starts its life as the two-dimensional
anomaly action for a gauge potential, $A_\mu$ \gaugeanom:
\eq
\Scl_{\rm anom} = \frac{1}{4 \pi} \; F^{\mu\nu} \; \left(
\frac{1}{\Square} \right) \; F_{\mn}.
\label\GAresult
\eeq
Eq.~\anomaction\ is obtained from this when $A_\mu$ is
restricted (in Minkowski signature)
to be transverse:
\eq
A_\mu = \hf \; \epsilon_\mn\,
\partial^\nu \, \rho .
\label\keysubst
\eeq

Now comes the key point. If the same replacement,
eq.~\keysubst, were made in {\it Euclidean} signature, as is done in refs.~\bg,
then one instead obtains a $\rho$ kinetic term having
the {\it opposite} sign. One obtains
opposite-sign actions depending on whether or not eq.~\keysubst\
is applied in Minkowski or Euclidean signature.

\ref\ostpos{
 K. Osterwalder and R. Schrader, \cmp{31}{73}{83},
{\it Helv. Phys. Acta}
{\bf 46} (1973) 277; \cmp{42}{75}{281}.}

Which sign is correct?  In a string-theory context, where the fundamental
path-integral formulation involves the Euclidean action, one is led to
the assignments of refs. ~\bg. On the other hand, if the fundamental
theory is defined as a Minkowski space field theory,
 it is the Minkowski
sign which is right. Euclidean conventions are generally defined
to reproduce Minkowski-space results. The correct substitution
which restricts $A_\mu$ to be transverse in Euclidean signature
is
\eq
A_m = {i\over 2} \; \epsilon_{mn} \, \partial^n \rho,
\label\euclsubst
\eeq
where the key difference from eq.~\keysubst\ is the factor of `$i$'.
Besides ensuring the equivalence of the Minkowski- and Euclidean-signature
anomaly actions, this factor of `$i$' is required for the unitarity
of the Euclidean action. That is, terms linear in $A_m$, such as
$i \ol\psi \gamma^m \psi \; A_m$, do not satisfy the
Osterwalder-Schraeder (OS) positivity condition \ostpos\ unless eq.~\euclsubst\
is used instead of eq.~\keysubst. (The OS condition is the
Euclidean equivalent of the Minkowski-signature
condition of the reality of the action, as required by unitarity. A similar
argument in four dimensions implies the standard result that
the $CP$-violating $\theta$-term of QCD has an imaginary coefficient
in Euclidean signature.)
We conclude that our eq.~\anomaction\ is the correct expression
for the super-Liouville action.
%

\listrefs

\vfill\eject

\bye